%% file: lat13-386.tex
\documentclass{PoS}

\usepackage{bm}
\usepackage{xifthen}

\newcommand{\text}[1]{\mathrm{#1}}
\newcommand{\arXiv}[2][hep-lat]{\href{http://arXiv.org/abs/#2}{#2%
\ifthenelse{\isempty{#1}}{}{ [#1]}}}

\title{Heavy-meson semileptonic decays for the Standard Model and beyond}

\ShortTitle{Heavy-meson semileptonic decays}

\vspace*{-6pt}
\author{
Yuzhi Liu,$^{a,b,c}$
Ran Zhou$,^{d,b}$
Jon A.~Bailey\thanks{Supported by the Ministry of Education under NRF Basic Science Grant No.\
2013009149.} ,$^{e}$ 
A.~Bazavov,$^{f,a}$ 
C.~Bernard,$^g$ 
C.~M.~Bouchard,$^h$
C.~DeTar,$^i$
Daping~Du,$^{j,k}$
A.~X.~El-Khadra,$^{j,b}$
J.~Foley,$^i$
E.~D.~Freeland,$^l$
E.~G\'amiz,$^m$
Steven~Gottlieb,$^d$
U.~M.~Heller,$^n$
R.~D.~Jain,$^j$
Jongjeong~Kim,$^o$
\speaker{A.~S.~Kronfeld},$^b$
J.~Laiho$^{p,k}$
L.~Levkova,$^i$
P.~B.~Mackenzie,$^b$
Y.~Meurice,$^a$
D.~Mohler,$^b$
E.~T.~Neil,$^{b,c,q}$
M.~B.~Oktay,$^i$
Si-Wei Qiu,$^i$
J.~N.~Simone,$^b$
R.~Sugar,$^r$
D.~Toussaint,$^o$ and
R.~S.~Van~de~Water$^b$ \\
$^a$Department of Physics and Astronomy, University of Iowa, Iowa City, Iowa, USA \\
$^b$Fermi National Accelerator Laboratory,\thanks{Operated by Fermi Research Alliance, LLC, under Contract
    No.~DE-AC02-07CH11359 with the US DOE.}~ Batavia, Illinois, USA \\
$^c$Department of Physics, University of Colorado, Boulder, Colorado, USA \\
$^d$Department of Physics, Indiana University, Bloomington, Indiana, USA \\
$^e$Department of Physics and Astronomy, Seoul National University, Seoul, South Korea \\
$^f$Physics Department, Brookhaven National Laboratory,\thanks{Operated by Brookhaven Science Associates, 
    LLC, under Contract No.\ DE-AC02-98CH10886 with the US DOE.}~ Upton, New~York, USA \\
$^g$Department of Physics, Washington University, St.~Louis, Missouri, USA \\
$^h$Department of Physics, The Ohio State University, Columbus, Ohio, USA \\
$^i$Department of Physics and Astronomy, University of Utah, Salt Lake City, Utah, USA \\
$^j$Physics Department, University of Illinois, Urbana, Illinois, USA \\
$^k$Department of Physics, Syracuse University, Syracuse, New~York, USA \\
$^l$	Liberal Arts Department, School of the Art Institute of Chicago, Chicago, Illinois, USA
$^m$CAFPE and Departamento de F\`isica Te\'orica y del Cosmos, Universidad de Granada, Granada, Spain \\
$^n$American Physical Society, Ridge, New York, USA \\
$^o$Department of Physics, University of Arizona, Tucson, Arizona, USA \\
$^p$SUPA, School of Physics \& Astronomy, University of Glasgow, Glasgow, UK \\
$^q$RIKEN-BNL Research Center, Brookhaven National Laboratory, Upton, New~York, USA \\
$^r$Department of Physics, University of California, Santa Barbara, California, USA \\
E-mail: \email{ask@fnal.gov}, \email{yuzhi.liu@colorado.edu}, \email{zhouran@fnal.gov}}

\author{Fermilab Lattice and MILC Collaborations}

\abstract{We calculate the form factors for the semileptonic decays $B_s\to K\ell\nu$ and $B\to K\ell\ell$
with lattice QCD.
We work at several lattice spacings and a range of light quark masses, using the MILC 2+1-flavor asqtad
ensembles.
We use the Fermilab method for the $b$ quark.
We obtain chiral-continuum extrapolations for $E_K$ up to $\sim1.2$~GeV and then extend to the entire
kinematic range with the model-independent $z$~expansion.
\vspace*{4pt}}

\FullConference{31st International Symposium on Lattice Field Theory - LATTICE 2013 \\
		July 29 - August 3, 2013\\
		Mainz, Germany}

\begin{document}

\section{Introduction}
\label{sec:intro}

In this report, we consider the form factors for two semileptonic processes: $B_s\to K\ell\nu$ and 
$B\to K\ell\ell$.
With lattice QCD, their analysis runs on parallel tracks, but their physics motivations differ.
The first is a $b\to u$ transition and can be used---with measurements from Belle~II or LHC$b$---to
determine the CKM matrix element~$|V_{ub}|$.
The other is a $b\to s$ transition; because the standard model (SM) amplitude is a one-loop penguin diagram,
it could be sensitive to non-SM physics, sought by Belle, BaBar, CDF, LHC$b$, and Belle~II
\cite{Zhou:2013uu}.

We compute the vector, scalar, and tensor form factors, defined by ($\bar{q}$ denotes $\bar{s}$, $\bar{d}$, 
or $\bar{u}$)
\begin{eqnarray} \hspace*{-2em}
    \langle K(k)|\bar{q}\gamma^\mu b|B_{(s)}(p)\rangle & = &
		\left[p^\mu+k^\mu-(M_{B_{(s)}}^2-M_K^2)\,q^{-2}\,q^\mu\right] f_+(q^2) +
		(M_{B_{(s)}}^2-M_K^2)\,q^{-2}\,q^\mu f_0(q^2) \nonumber \\
        & = & \sqrt{2M_{B_{(s)}}}\left[v^\mu f_\parallel(E_K)
            + k_\perp^\mu f_\perp(E_K) \right], 
\end{eqnarray}
\vspace*{-24pt}
\begin{eqnarray}
	\langle K(k)|\bar{q} b|B_{(s)}(p)\rangle & = &
        f_0(q^2) \left(M_{B_{(s)}}^2-M_K^2\right)/(m_b-m_q), \\
	\langle K(k)|\bar{q}i\sigma^{\mu\nu} b|B_{(s)}(p)\rangle & = & 2  f_T(q^2)
		\left(p^\mu k^\nu-p^\nu k^\mu\right)/\left(M_{B_{(s)}}+M_K\right), 
\end{eqnarray}
where $q=p-k$, $v^\mu = p^\mu/M_{B_{(s)}}$, $k_\perp^\mu=k^\mu-(v\cdot k)v^\mu$, and the kaon energy $E_K$ is
related to~$q$ via $q^2=M_{B_{(s)}}^2 + M_K^2-2M_{B_{(s)}}E_K$.
In the limit of a very heavy $B_{(s)}$ meson,
$f_T(q^2)=(M_{B_{(s)}}+M_K)(2M_{B_{(s)}})^{-1/2}f_\perp(E_K)+\mathrm{O}(1/m_b)$.
We obtain the matrix elements from three-point correlation functions and the energies and amputation factors
from two-point correlation functions~\cite{Bailey:2008wp,Zhou:2011be,Zhou:2012dm,Du:2013kea}.

The ensembles of lattice gauge fields are listed in Table~\ref{tab:ens}.
They employ an improved gluon action and the asqtad action for $2+1$ flavors of sea quark.
For the valence quarks, we again use the asqtad action for the light quarks, and the Fermilab method for the 
heavy $b$ quark.
We match the lattice currents to the continuum with the mostly nonperturbative method of 
Ref.~\cite{ElKhadra:2001rv}.
We have used our experience with $B\to K\ell\ell$ to guide the run plan for $B_s\to K\ell\nu$.
The HPQCD Collaboration has used these ensembles for $B\to K\ell\ell$~\cite{Bouchard:2013eph}, with 
special emphasis on phenomenology~\cite{Bouchard:2013mia}.

Our two analyses share many common features.
In the past~\cite{Zhou:2011be,Zhou:2012dm}, we reported on the correlator fits for $B\to K\ell\ell$.
In Sec.~\ref{sec:Bs}, we outline the analogous steps for our analysis of $B_s\to K\ell\nu$.
Then we discuss the chiral-continuum extrapolation of $B\to K\ell\ell$ in Sec.~\ref{sec:B}.
Future work on this aspect of the $B_s\to K\ell\nu$ analysis will follow a similar strategy.
Some outlook is given in Sec.~\ref{sec:end}.

\begin{table}[t]
    \centering
    \caption{Ensembles used in the FCNC $B\to K\ell\ell$ project, with valence strange mass equal to the sea 
        strange mass.
        The CKM $B_s\to K\ell\nu$ project used a valence strange mass closer to the physical value, on the 
        subset with an entry for $am_s^{\mathrm{val}}$ $(B_s)$.
        A~full description of these ensembles can be found in Ref.~\cite{Bazavov:2009bb}.}
    \label{tab:ens}
    \newcommand{\h}{\hphantom{0}}
    \begin{tabular}{ccr@{/}lrccc}
        \hline\hline
        $a$~(fm) & $N_S^3\times N_T$ & $am_l$&$am_s$ & \# confs & \# sources & $am_s^{\mathrm{val}}$ $(B_s)$ & 
            $am_s^{\mathrm{val}}$ $(B)$ \\
        \hline
        $\approx 0.12$ & $20^3\times 64$ &  0.01&0.05    & 2259~ & 4 &        & 0.05\h\h \\
        $\approx 0.12$ & $20^3\times 64$ & 0.007&0.05    & 2110~ & 4 &        & 0.05\h\h \\
        $\approx 0.12$ & $20^3\times 64$ & 0.005&0.05    & 2099~ & 4 & 0.0336 & 0.05\h\h \\
        $\approx 0.09$ & $28^3\times 96$ & 0.0124&0.031  & 1996~ & 4 &        & 0.031\h \\
        $\approx 0.09$ & $28^3\times 96$ & 0.0062&0.031  & 1931~ & 4 & 0.0247 & 0.031\h \\
        $\approx 0.09$ & $32^3\times 96$ & 0.00465&0.031 &  984~ & $4_{\rm FCNC}\quad 8_{\rm CKM}$ & 0.0247 &
            0.031\h \\
        $\approx 0.09$ & $40^3\times 96$ & 0.0031&0.031  & 1015~ & $4_{\rm FCNC}\quad 8_{\rm CKM}$ & 0.0247 &
            0.031\h \\
        $\approx 0.09$ & $64^3\times 96$ & 0.00155&0.031 &  791~ & 4 & 0.0247 & 0.031\h \\
        $\approx 0.06$ & $48^3\times144$ & 0.0036&0.018  &  673~ & 4 &        & 0.0188 \\
        $\approx 0.06$ & $64^3\times144$ & 0.0018&0.018  &  673~ & 4 & 0.0177 & 0.0188 \\
        \hline\hline
    \end{tabular}
    \vspace*{-12pt}
\end{table}

\input ckm

\input fcnc

\section{Outlook}
\label{sec:end}

\vspace*{-1pt}
Our nearly final results for $B\to K\ell\ell$ will have errors roughly of 3--8\%, for 
$q^2>17~\mathrm{GeV}^2$.
For smaller $q^2$, the form factors become small and inevitably have large relative error.
These results provides a reasonable forecast of the error budget for $B_s\to K\ell\nu$, once that project is 
complete.

\acknowledgments
\vspace*{-1pt}
This work was supported by the U.S. Department of Energy, the National Science Foundation, and the URA 
Visiting Scholars' Program.
Computations were carried out at the Argonne Leadership Computing Facility,
the National Center for Atmospheric Research,
the National Center for Supercomputing Resources,
the National Energy Resources Supercomputing Center,
the National Institute for Computational Sciences, 
the Texas Advanced Computing Center, and
the USQCD facilities at Fermilab, under grants from the NSF and DOE.

\end{document}

%% file: ckm.tex
\section{CKM Mode $\bm{B_s\to K\ell\nu}$}
\label{sec:Bs}

To obtain kaon energies, the $B_s$ mass, and amputation factors, we fit the meson two-point data to the 
functional form 
\begin{equation}
    C_{ss'}(t) = 
        \sum_{n=0}^{N-1} \left[ A_{n,s} A_{n,s'} \left(e^{-E_n t} + e^{-E_n (N_T-t)} \right) 
            - (-1)^t A_{n,s}' A_{n,s'}' \left(e^{-E_n' t} + e^{-E_n' (N_T-t)} \right)\right],
    \label{eq:BstoK:2pts}
\end{equation}
where the subscripts $s$ and $s'$ denote the source and sink, 
the coefficients $A_{n,s}$ denote the projection of source or sink $s$ onto the $n$th excitation, and
$N$ is the number of states retained in the fits.
The second tower of states arises because with staggered fermions parity is nonlocal in time.
We use a local source and sink for the kaon, and a $1S$-smeared source and both local and
$1S$-smeared sinks for the $B_s$ meson.
This combination has worked well in our $B\to \pi$~\cite{Bailey:2008wp,Du:2013kea} and 
$B\to K$~\cite{Zhou:2011be,Zhou:2012dm} projects.

\begin{figure}[bp]
    \centering
    \includegraphics[width=0.435\textwidth]{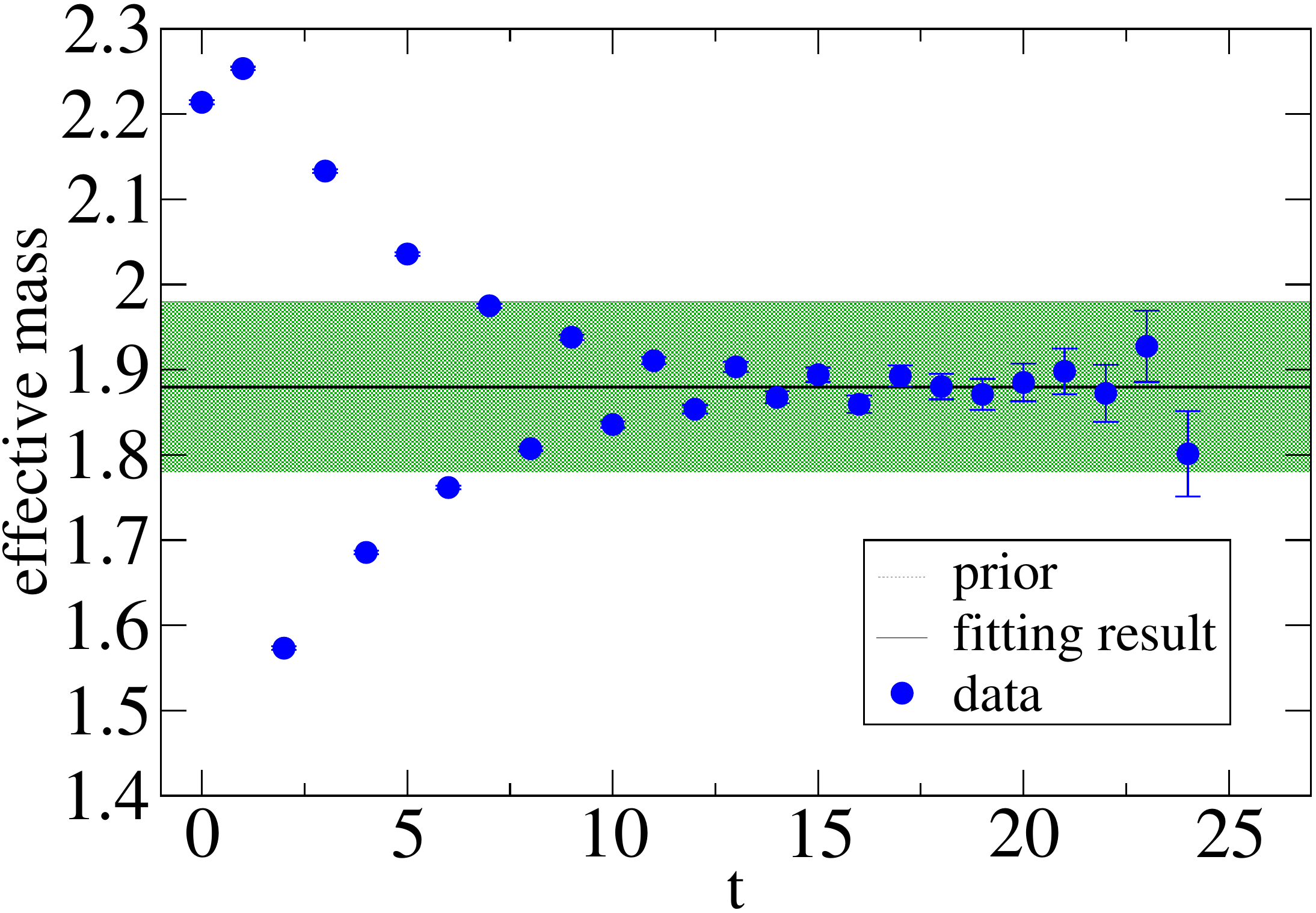}\hfill
    \includegraphics[width=0.525\textwidth]{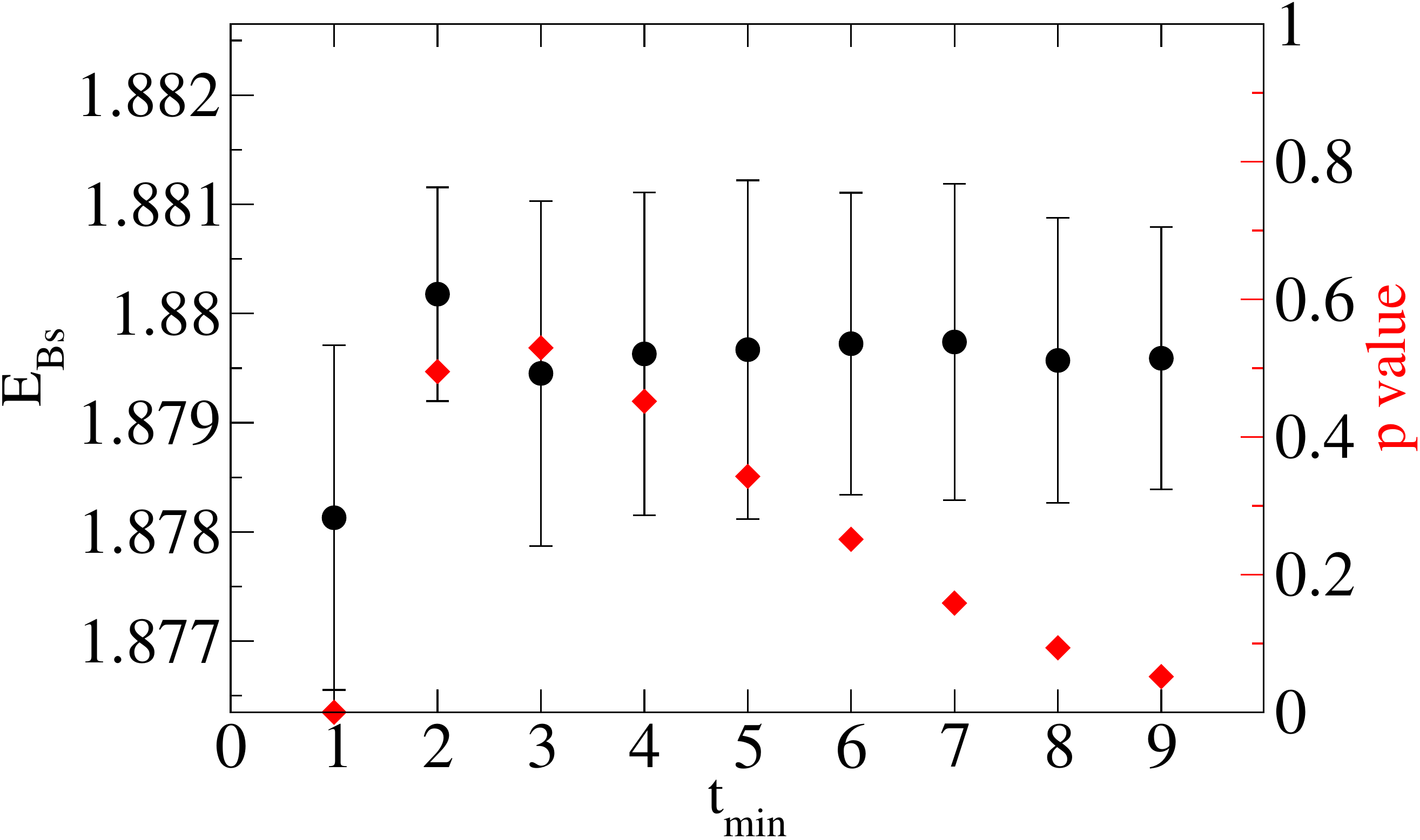}
    \caption{$B_s$ two-point correlator with $\bm{k}=\bm{0}$ for the $a\approx0.12$~fm, 
        $am_l/am_s=0.005/0.05$ ensemble.  
        The effective mass plot (left plot) leads us to choose the prior shown in green.
        In the same vein, we use scaled correlators $C_{ss'}(t)e^{M_{\mathrm{eff}}t}$ to set priors on 
        $A_{n,s^{(\prime)}}$. 
        The fitted $B_s$ mass (right plot, left axis) and $p$~value (right plot, right axis) 
        lead us to choose $t_\mathrm{min}=3$.
        The $p$-value calculation is explained in the text.}  
    \label{fig:Bs_effmass_tmin}
\end{figure}

We fix the number of states $N=3$ for all correlator fits.
Figure~\ref{fig:Bs_effmass_tmin} shows an example of an effective mass plot and a stability plot for the
$B_s$ meson, on the $a\approx0.12$~fm, $am_l/am_s=0.005/0.05$ ensemble.
We use these plots to choose fit intervals $[t_\mathrm{min},t_\mathrm{max}]$ and to set loose Bayesian priors
for the ground states.
We take priors of moderate width for excited states.
The goodness of fit $p$ value is calculated from the degrees of freedom and the augmented $\chi^2$, which
includes contributions from the priors.
With our moderate priors, the effective count of degrees of freedom lies between the usual count (\# of data
points minus \# of parameters) and the augmented one (\# of data points).
We use the former in Fig.~\ref{fig:Bs_effmass_tmin}, so the plotted $p$~values are underestimates.

To obtain the semileptonic matrix elements, we fit the three-point data to the functional form
\begin{equation}
    \vspace*{-12pt}
    C_{ss'}^\mu(t,T) = \sum_{m,n=0}^{N-1} (-1)^{mt} (-1)^{n(T - t)} 
        A_{m,s}^K V^\mu_{mn} A_{n,s'}^{B_s} e^{-E_{K}^m t} e^{-M_{B_s}^n (T - t)},
    \vspace*{4pt}
    \label{eq:BstoK:3pts}
\end{equation}
where 
$V^\mu_{mn}=\langle K_m|V^\mu|B_{s,n}\rangle/\left(2 E_K^m\,2M_{B_s}^n\right)^{1/2}$ yields the
form factors. 
We put the kaon at several source times (``\# sources'' in Table~\ref{tab:ens}) and choose two different
locations for the $B_s$-meson sink, $T=T_\mathrm{sink}$ and $T_\mathrm{sink}+1$.
We perform a combined fit to the two-point and three-point correlation functions.
The three-point fitting ranges are chosen to be between $t_\mathrm{min}^K$ and $T-t_\mathrm{min}^{B_s}$,
with $T=T_\mathrm{sink}$ or $T_\mathrm{sink}+1$.
To set the prior for $V^\mu_{00}$, we follow a strategy like that explained with Fig.~\ref{fig:Bs_effmass_tmin}.

\begin{figure}[h]
    \centering
    \includegraphics[width=0.5\textwidth]{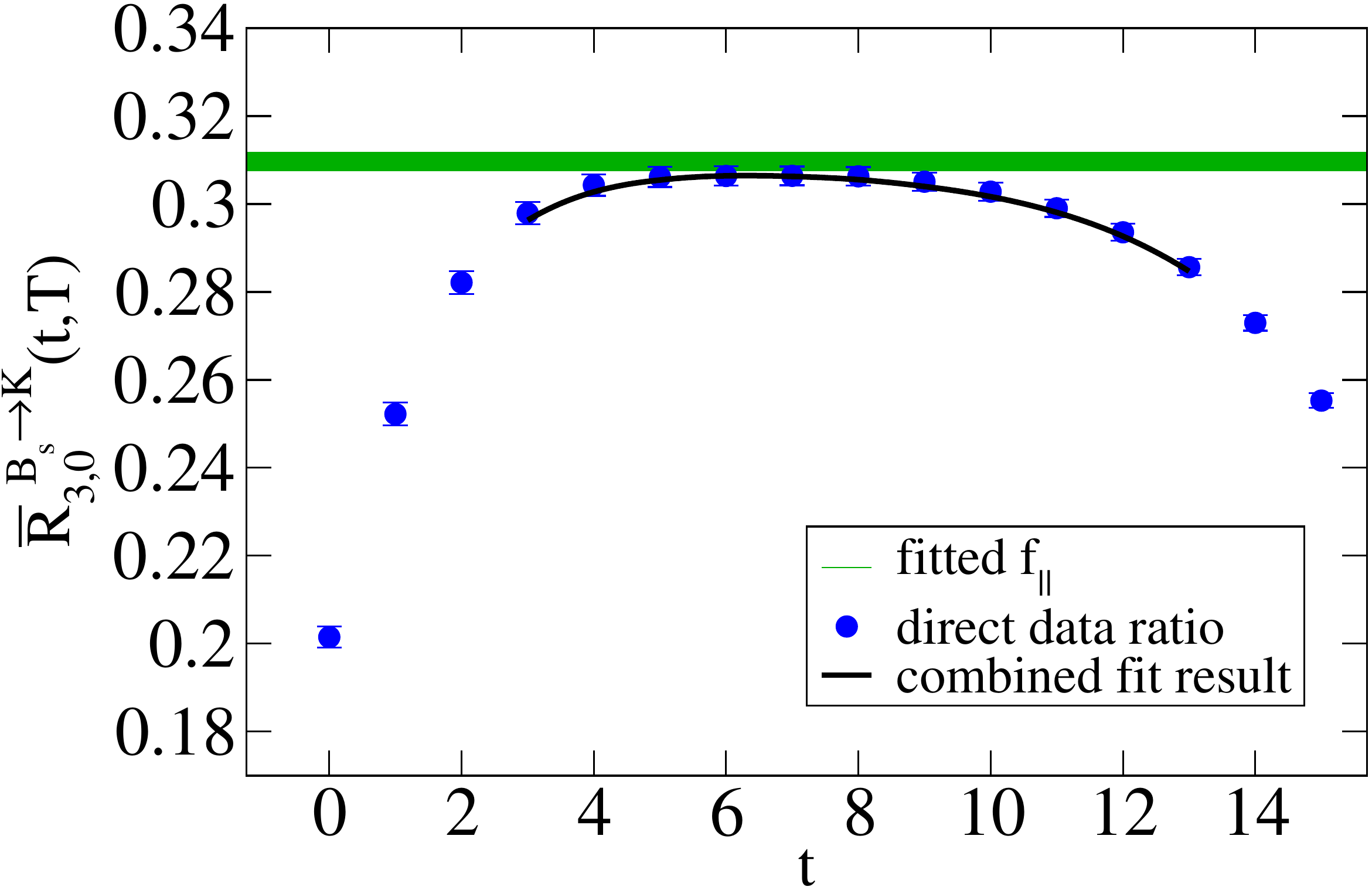}
    \caption{The form factor from different fits.
        The green band is the combined fit discussed in the text.
        The blue points are the ratio used in Refs.~\cite{Bailey:2008wp} and~\cite{Zhou:2012dm}; 
        the plateau lies a bit lower.
        The black curve shows the combined fit's reconstruction of the ratio, showing that the small 
        difference comes from excited states.}
    \label{fig:3pt_fit}
\end{figure}

In Refs.~\cite{Bailey:2008wp} and~\cite{Zhou:2012dm}, we used a plateau fit to a ratio of two- and 
three-point functions.
Here, we carry out a combined fit, so that we can easily take the excited states' contributions into account.
Figure~\ref{fig:3pt_fit} shows that the fitted $V^\mu_{00}$ lies above the ratio $\bar{R}$ constructed from 
the two- and three-point data,
again on the $a\approx0.12$~fm, $am_l/am_s=0.005/0.05$ ensemble, with $\bm{k}=\bm{0}$.
The difference stems from the excited state contribution.
Ignoring it yields a value for $V^\mu_{00}$ with a small bias, which with our current statistics is 
significant in some cases, such as the one shown in Fig.~\ref{fig:3pt_fit}. 

We have completed these fits for all ensembles indicated in Table~\ref{tab:ens}.
Since the conference, we have begun to carry out the chiral-continuum extrapolation.
Our approach will follow that of our $B \to K\ell\ell$ work, which is discussed in the next section.

%% file: fcnc.tex
\section{FCNC Mode $\bm{B\to K\ell\ell}$}
\label{sec:B}

We calculate $B\to K\ell\ell$ form factors from the ratio of three-point and two-point correlation functions.
In general, $B\to K\ell\ell$ has larger errors than $B_s\to K\ell\nu$.
In this case, we find that the plateau fit is consistent, within errors, with the combined fit of
Sec.~\ref{sec:Bs}.
We choose the simpler fit for this analysis.
To obtain physical results, we perform a chiral-continuum extrapolation using heavy-light meson staggered
chiral perturbation theory (HMS$\chi$PT) \cite{Aubin:2007mc}.
We have found that next-to-next-to-leading order SU(3) HMS$\chi$PT does not describe the $f_\parallel$ data
well~\cite{Zhou:2011be,Zhou:2012dm}.
The fits have low $p$~values, and the curves behave unphysically for $E_K$ outside the range where we have
data.
On~the other hand, SU(2) HMS$\chi$PT \cite{Bijnens:2010ws} describes the data well, even at next-to-leading
order (NLO).

\begin{figure}[bp]
    \includegraphics[scale=0.5]{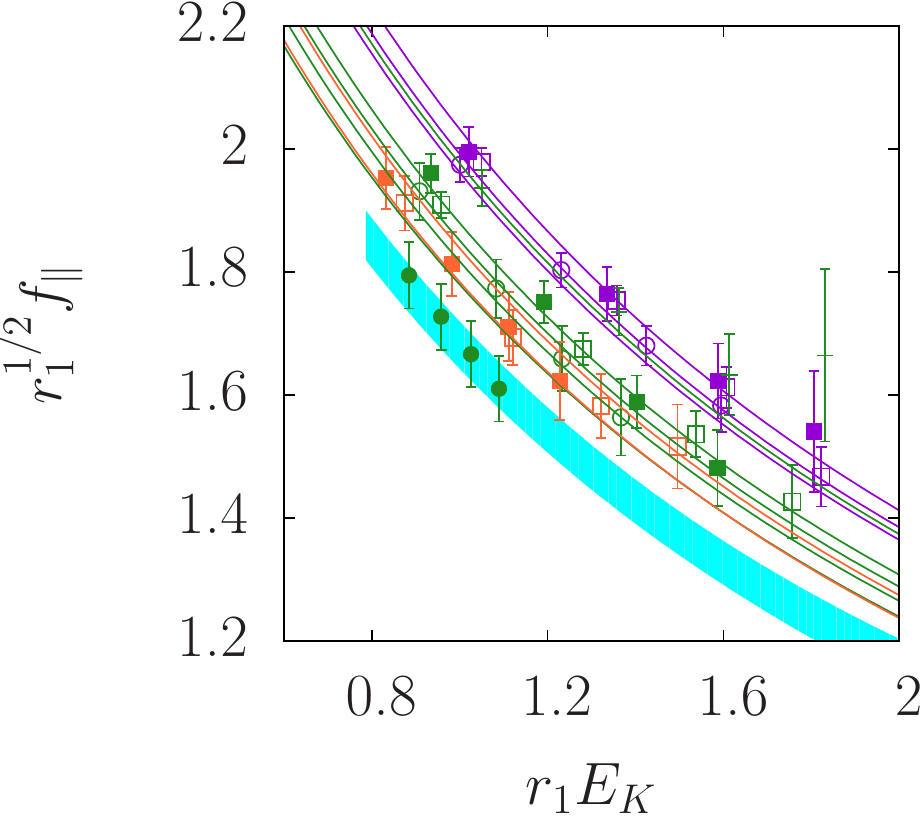} \hfill
    \includegraphics[scale=0.5]{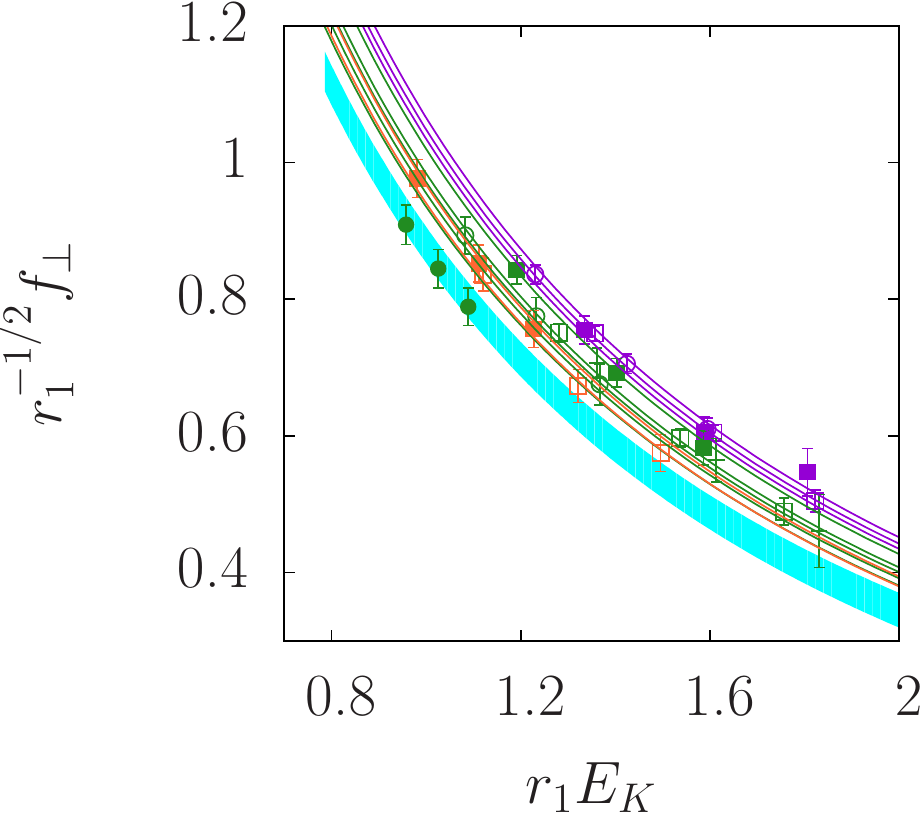} \hfill
    \includegraphics[scale=0.5]{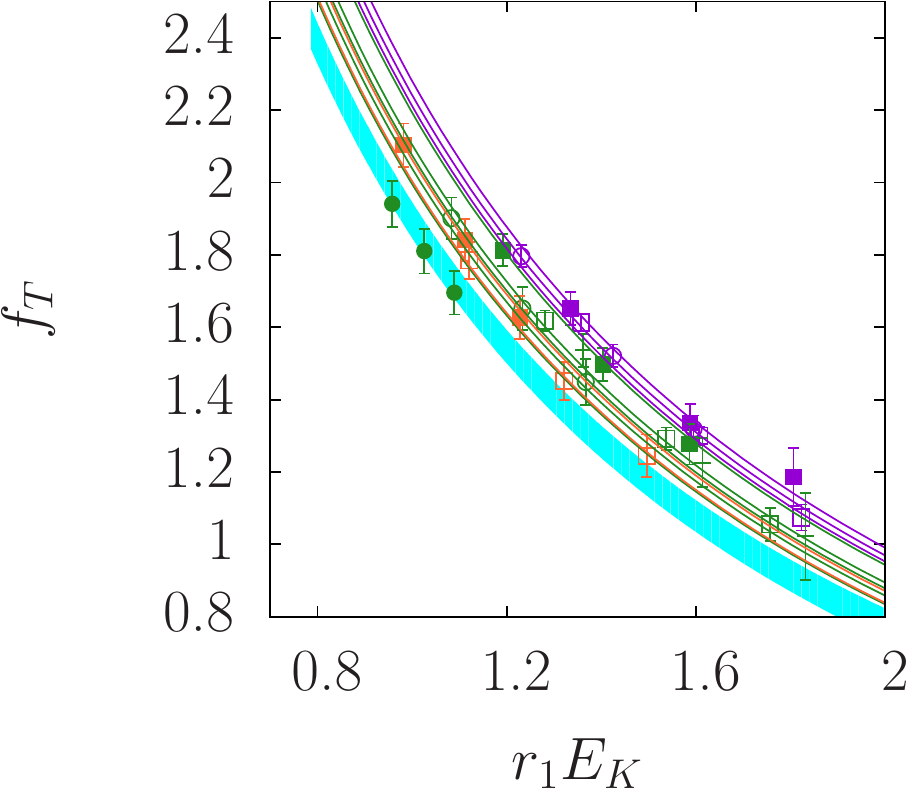}
    \caption{Chiral-continuum extrapolations of (left to right) $f_\parallel$, $f_\perp$, and $f_T$ with NLO
        SU(2) HMS$\chi$PT.
        Bars, unfilled squares, filled squares, unfilled circles, and filled circles denote 
        $m_l/m_s = 0.4$, 0.2, 0.14, 0.1, and 0.05 data, respectively.
        Violet, green, and orange lines represent the fit curves on $a\approx0.12$, 0.09, and 0.06~fm
        ensembles, and the cyan band represents the continuum extrapolated curve and its error.}
    \label{fig:combined_su2_fpara_fperp_ft_chpt_fit}
\end{figure}

For our main fit, we consider the functional form,
\begin{equation}
        f_{\rm {pole}} = \frac{C^{(0)}}{f_\pi(E_K + \Delta_{B_s^*})} \left[ 1 + {\rm logs} + 
        C^{(1)} \chi_l^{\rm val} + C^{(2)} \chi_s^{\rm val} + C^{(3)} \chi_E + C^{(4)} \chi_{a^2} \right] ,
    \label{eq:chiral.pole}
\end{equation}
where the dimensionless quantities $\chi_i$ are defined in Ref.~\cite{Bailey:2008wp}.
We include the strange-quark mass related term $C^{(2)} \chi_s^{\rm val}$ to account 
for $m_s$ dependence in~$C^{(0)}$. 
Equation~(\ref{eq:chiral.pole}) builds in a pole in $E_K$ at $-\Delta_{B_s^*}$, where
$\Delta_{B_s^*} = M_{B_s^*}-M_B$.
For~$f_\perp$ and~$f_T$, we use the lowest-lying, $J^P=1^-$ pole and set $\Delta_{B_s^*}=0.1358$~GeV.
For $f_\parallel$, one should look at structures in the $J^P=0^+$ channel for $E_K<0$ or, equivalently,
$q^2>(M_{B_{(s)}}+M_K)^2$.
Now no single state clearly dominates the physical region.
Without attaching a deep meaning to it, we simply incorporate one pole into the fit, with a prior of central
value $\Delta_{B_s^*}=0.44$~GeV~\cite{Bardeen:2003kt} and width~0.50~GeV.
We impose Gaussian priors with central values~0 and width~2 on the NLO parameters~$C^{(i)}$.
The chiral logarithms in Eq.~(\ref{eq:chiral.pole}) depend on the $B$-$B^*$-$\pi$ coupling, $g_\pi$, which
we constrain with a prior of central value 0.45 and width 0.08, based on recent direct 
calculations~\cite{Bulava:2011yz,Detmold:2011bp}. 
Finally, we add further terms to Eq.~(\ref{eq:chiral.pole}) to describe heavy-quark discretization effects,
in the manner used for heavy-light decay constants~\cite{Bazavov:2011aa}.
The data and chiral-continuum fits are shown in Fig.~\ref{fig:combined_su2_fpara_fperp_ft_chpt_fit}.

\begin{figure}[bp]
    \includegraphics[scale=0.45]{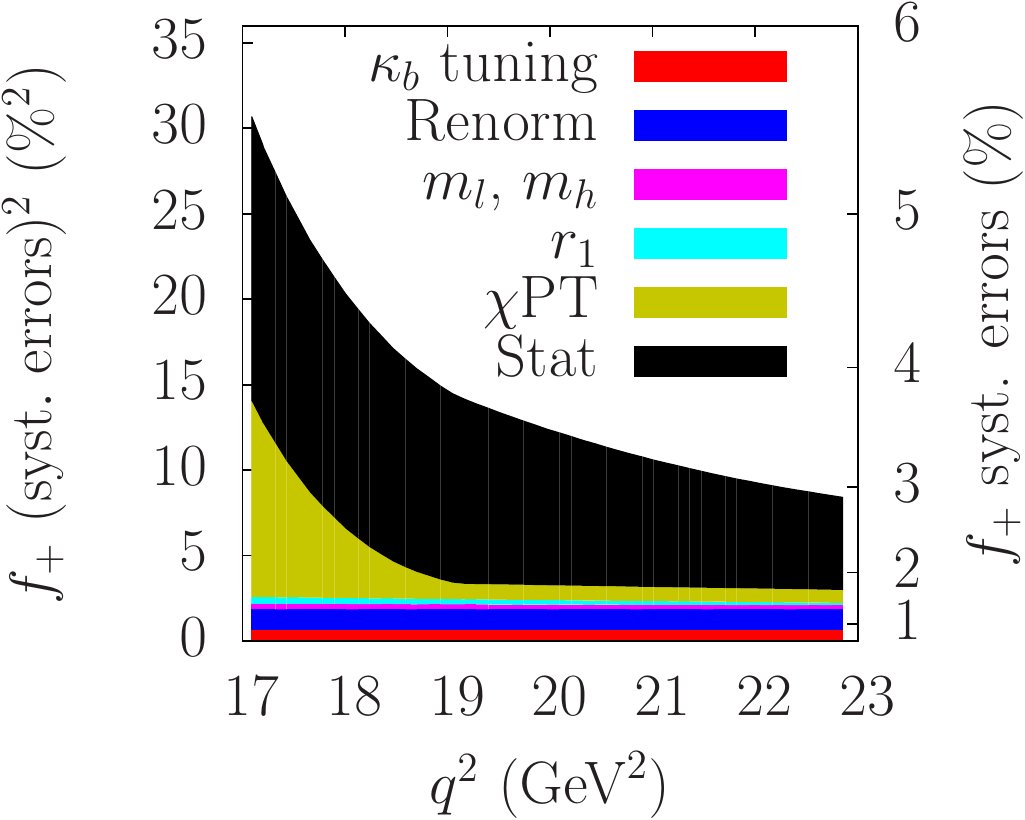} \hfill
    \includegraphics[scale=0.45]{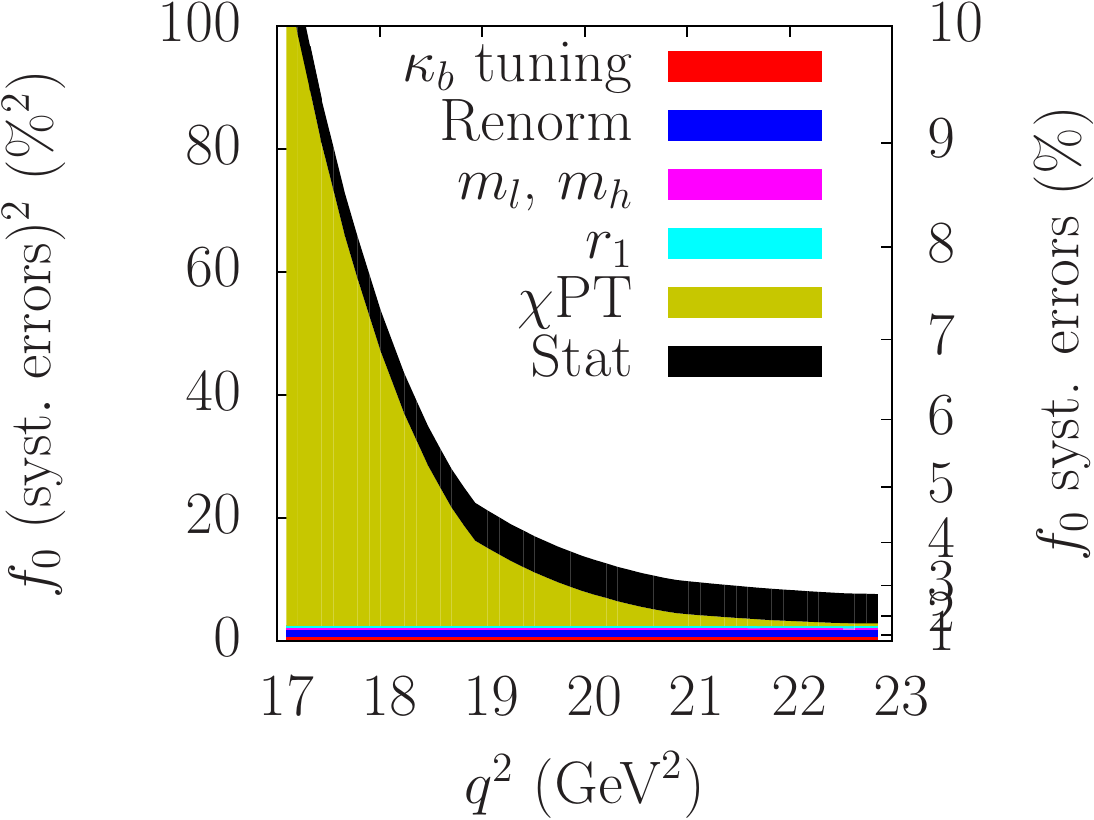} \hfill
    \includegraphics[scale=0.45]{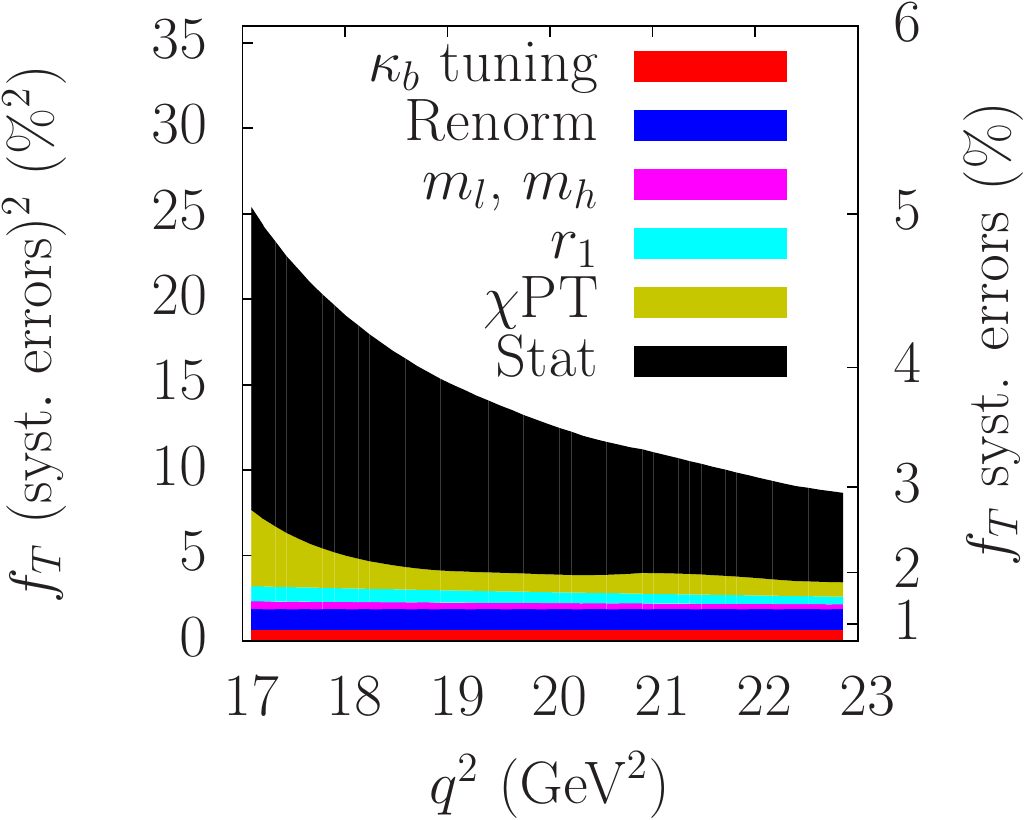} \hfill
    \caption{Statistical and systematic error contributions to (left to right) $f_+$, $f_0$, and $f_T$.
        The left vertical axis shows the squares of the errors added in quadrature, while the right vertical 
        axis shows the errors themselves.}
    \label{fig:fp_f0_ft_errors}
\end{figure}

Our chiral-continuum extrapolation incorporates statistical errors, an uncertainty in $g_\pi$, and
heavy- and light-quark discretization effects.
We must, however, consider additional sources of uncertainty, stemming from the choice of fitting Ansatz, as
well as from heavy-light current renormalization, lattice-scale determination, light-, strange-, and
heavy-quark mass tuning, and finite volume effects.
For example, we have also removed the pole form from the $f_\parallel$ fits, modeling the curvature in $E_K$
with an extra term in the series inside the bracket.
Our estimates for the statistical and systematic errors are shown, as a function of $q^2$ in
Fig.~\ref{fig:fp_f0_ft_errors}.
%

Our chiral-continuum extrapolation results only cover a limited range of $E_K$ (or $q^2$), for several 
reasons.
In particular, the HMS$\chi$PT description breaks down for large $E_K$.
In order to extend the form factors to higher kaon energy, i.e., to $q^2=0$, we use the model-independent
$z$~expansion.
This method maps $q^2$ to a new variable $z$ such that $|z|<1$,
\begin{equation}
    z(q^2,t_0) =\frac{\sqrt{t_+-q^2}-\sqrt{t_+-t_0}}{\sqrt{t_+-q^2}+\sqrt{t_+-t_0}} ,
\end{equation}
where $t_\pm=(M_B\pm M_K)^2$, and we take $t_0=(M_B+M_K)(\sqrt{M_B}-\sqrt{M_K})^2$~\cite{Bourrely:2008za}.
The physical region of $q^2$ is mapped onto the real interval $|z|<0.16$.
We apply this mapping to the output of the chiral-continuum extrapolation 
(Fig.~\ref{fig:combined_su2_fpara_fperp_ft_chpt_fit}) with the full error budget 
(Fig.~\ref{fig:fp_f0_ft_errors}).

With the new variable $z$, we expand the form factors as~\cite{Bourrely:2008za} 
\begin{eqnarray}
    f_{+,T}(q^2) &=& \frac{1}{1-q^2/M_{B_s^*}^2}\sum_{k=0}^{K-1} b_k \left[ z^k -(-1)^{k-K}\frac{k}{K}z^K \right], 
    \label{eq:f+zexp} \\
    f_0(q^2) &=& \frac{1}{1-q^2/M_{B_s^*}^2}\sum_{k=0}^{K-1} b_k  z^k, 
    \label{eq:f0zexp} 
\end{eqnarray}
where $M_{B_s^*}^2$ accounts for the location of the pole in the form factors.
We use here the same pole positions as in the chiral-continuum fit.
To obtain the $b_k$, we generate five synthetic data points at $q^2 = (22.86, 21.59, 20.21, 18.70,
16.86)~\mathrm{GeV}^2$.
Because Eq.~(\ref{eq:chiral.pole}) has only two $E_K$-dependent pieces of information, we impose an SVD cut
when inverting the covariance matrix.
We fit $f_+$ and $f_0$, which are linear combinations of $f_\perp$ and $f_\parallel$, together, keeping the
four highest eigenmodes total.
For $f_T$ we keep two.
Fits to Eqs.~(\ref{eq:f+zexp}) and~(\ref{eq:f0zexp}) with $K=2$ or~3 are good and also naturally preserve 
the kinematic condition $f_+(0)=f_0(0)$,  provided we include the scalar pole in $f_0$.

\begin{figure}[bp]
    \centering
    \includegraphics[scale=0.5]{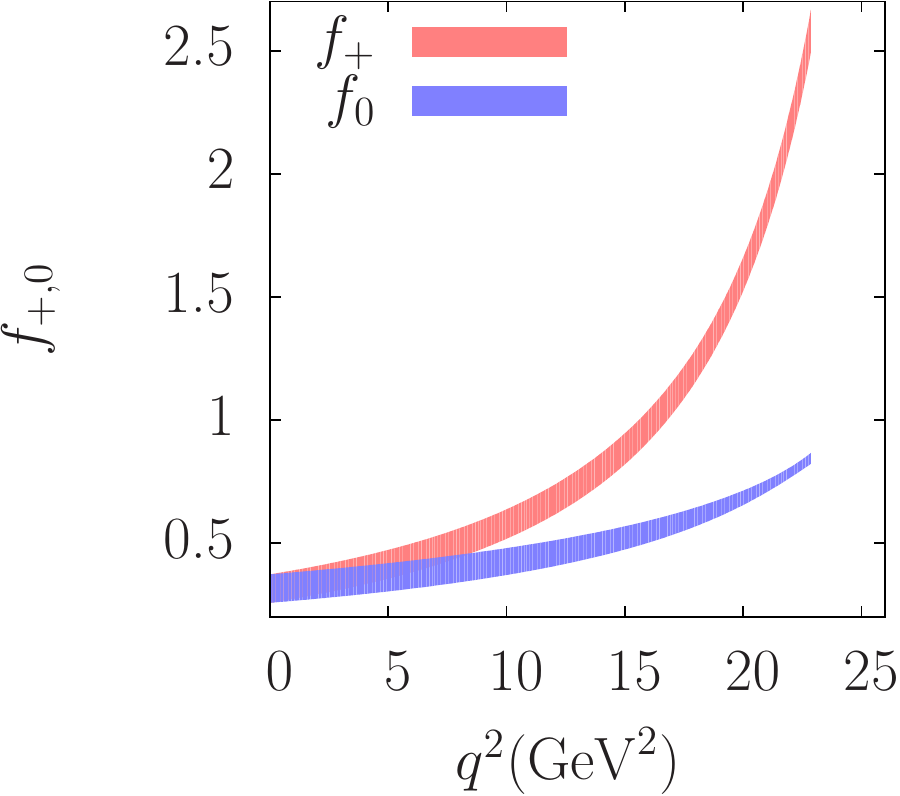}\quad\quad
    \includegraphics[scale=0.5]{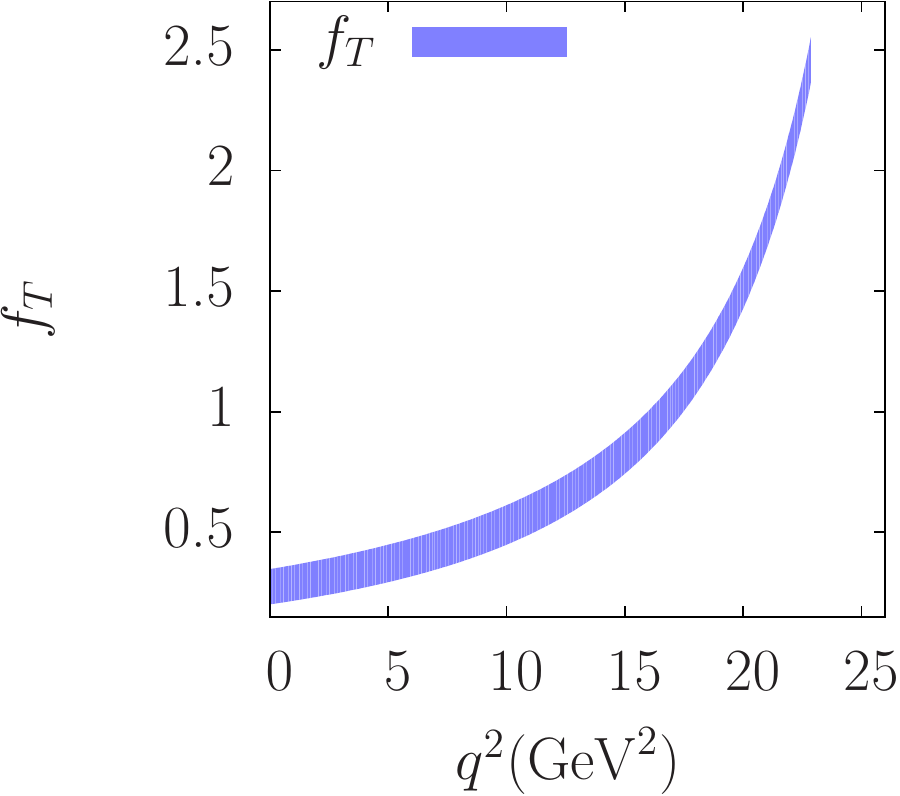}
    \caption{Final $z$-expansion fit of $f_+$ \& $f_0$ (left) and $f_T$ (right) \emph{vs.} $q^2$ with full 
        error band.
        Here we use three $b_j$ for each form factor and impose the kinematic condition $f_+(0)=f_0(0)$.}
    \label{fig:zexp_fvsq2}
\end{figure}

Having seen that the kinematic condition arises naturally, our central fits impose it as a constraint.
This step gives better control of the extrapolation error at low~$q^2$.
Unitarity, analyticity, and heavy-quark physics place upper bounds on $\sum_{jk}b_jB_{jk}b_k$, where $B_{jk}$
is a matrix explained in Ref.~\cite{Bourrely:2008za}.
The heavy-quark bound is more restrictive but only semiquantitative~\cite{Becher:2005bg}.
We put a prior on $\sum_{jk}b_jB_{jk}b_k$ of central value zero and width of 0.1 for $f_+$ and $f_T$, and
width 0.3 for $f_0$, which correspond to conservative interpretations of the heavy-quark bound.
Figure~\ref{fig:zexp_fvsq2} shows the outcome of these fits.
These results are nearly final, and we plan to publish the $z$-expansion coefficients with their errors and
correlations, so that the final output of our analysis can be used elsewhere.
\vspace*{-2pt}


